\documentclass[aps,pre,reprint,superscriptaddress,10pt]{revtex4-2}
\usepackage{amsmath}
\usepackage{amssymb}
\usepackage{graphicx}% Include figure files
\usepackage{dcolumn}% Align table columns on decimal point
\usepackage{bm}% bold math
\usepackage{hyperref}
\usepackage{courier}
\usepackage{braket}
\usepackage{float}
\usepackage[usenames,dvipsnames]{xcolor}
\hypersetup{
  colorlinks   = true,	%Colours links instead of ugly boxes
  urlcolor     = blue,	%Colour for external hyperlinks
  linkcolor    = blue,	%Colour of internal links
  citecolor   = blue	%Colour of citations#NavyBlue
}
\begin{document}
\newcommand{\ve}{\varepsilon}
\title{Attractive-repulsive interaction in coupled quantum oscillators}

\author{Bulti Paul}
\affiliation{Chaos and Complex Systems Research Laboratory, Department of Physics, University of Burdwan, Burdwan 713 104, West Bengal, India}
\author{Biswabibek Bandyopadhyay}
\affiliation{Department of Physics \& Astronomy, Northwestern University, 2145 Sheridan Rd. Evanston, IL 60208 USA}
\author{Tanmoy Banerjee}
\email[]{tbanerjee@phys.buruniv.ac.in}
%%\homepage[]{Your web page}
\thanks{he/him/his}
%%\altaffiliation{}
\affiliation{Chaos and Complex Systems Research Laboratory, Department of Physics, University of Burdwan, Burdwan 713 104, West Bengal, India}

\date{\today}

\begin{abstract}
We study the emergent dynamics of quantum self-sustained oscillators induced by the simultaneous presence of attraction and repulsion in the coupling path. We consider quantum Stuart-Landau oscillators under attractive-repulsive coupling and construct the corresponding quantum master equation in the Lindblad form. We discover an interesting symmetry-breaking transition from quantum limit cycle oscillation to quantum inhomogeneous steady state; 
This transition is contrary to the previously known symmetry-breaking transition from quantum homogeneous to inhomogeneous steady state. 
The result is supported by the analysis on the noisy classical model of the quantum system in the weak quantum regime. Remarkably, we find the generation of entanglement associated with the symmetry-breaking transition that has no analogue in the classical domain. This study will enrich our understanding of the collective behaviors shown by coupled oscillators in the quantum domain.  
\end{abstract}

\maketitle

%%%%%%%%%% Introduction ##########################

\section{Introduction}
\label{sec_intro}
Studies on coupled oscillators constitute a paradigmatic framework to understand diverse complex processes in the field of physics, biology, and engineering \cite{sync}. Several cooperative behaviors have been observed in coupled oscillators such as synchronization, chimeras, and oscillation quenching that enriched our understanding of the nature \cite{sync-boca}.
It is found that coupling function plays a pivotal role in shaping the observed emergent dynamics \cite{coupling-function-aneta1,coupling-function-aneta2}. A plethora of coupling schemes were studied in the literature and based on the synchronizing or desynchronizing effect of the coupling function they can be categorized into two broad types, namely attractive and repulsive coupling, respectively: while an attractive coupling tends to synchronize the oscillators, a repulsive coupling tends to desynchronize them \cite{majhi2020perspective}. 

However, real world systems are found to be governed by the simultaneous presence of attractive and repulsive coupling (we denote it by attractive-repulsive coupling). Examples are abound in the field of physics \cite{network-science}, ecology \cite{eco} and neuronal systems \cite{neuro-attrep} (see \cite{majhi2020perspective} for a recent review).
%%%%%%%%%%%
Chen et al. \cite{chen2009dynamics} investigated the dynamics of two coupled oscillators under attractive-repulsive coupling and found a rich variety of collective dynamics including amplitude death. 
\citet{strog-contra} studied the role of attractive-repulsive coupling in a network of coupled phase oscillators and found synchronized states among the contrarian (oscillators with repulsive coupling) and conformists (oscillators with attractive coupling).
\citet{hens} explored the occurrence of cessation of oscillation in the presence of additional repulsive link in a system of coupled oscillators.
The effect of attractive-repulsive coupling in a network of non-locally coupled oscillators were studied by \citet{laks-attrep}, and chimera death and oscillation death states were reported.   
Zhao et al. \cite{zhao2018amplitude} showed that uniform and mixed attractive-repulsive coupling among three coupled Stuart-Landau oscillators can exhibit diverse emergent behaviors. 
Several variants of attractive-repulsive coupling were studied in Refs.~\cite{dixit2019dynamics,dixit2020static} and transitions from oscillatory to steady states were found. Recently, the notion of attractive-repulsive coupling has been explored in swarmalator model and splaylike states are observed \cite{swarm}. 
%%%%%%%%%%

All the above studies were carried out for classical oscillators in the classical domain. The role of attractive-repulsive coupling on the emergent behavior of quantum oscillators  has not been studied yet: {\it In this work, for the first time, we study the attractive-repulsive coupling in coupled quantum oscillators in the quantum regime.} 
Our study is motivated by the fact that most of the well known collective behaviors shown by classical oscillators  are manifested in a counter intuitive way in the quantum domain due to the presence of inherent quantum mechanical constraints \cite{chia}. Studies on the quantum version of synchronization \cite{lee_prl,brud_prl1,brud-ann15,squeezing,blockade,brud-poch,senthil-quantum,chia-23}, oscillation suppression \cite{qad1,qad2}, and symmetry-breaking states \cite{qchm,qmod,qrev,qturing,qkerr,kato,qaging} indeed established several counter intuitive results that have no analogue in the classical domain. 
Further, the notion of attractive-repulsive interaction has been a vibrant area of research in the quantum science and engineering: this interaction has been investigated in the context of Fermi polaron \cite{polaron-prx}, dynamical band flipping in a system of correlated lattice fermions \cite{att-rep-prl}, quantum optical system with strongly interacting photon \cite{cantu2020repulsive}, photonic Bose–Hubbard dimer \cite{li2017mapping} to name a few. However, these studies have not considered self-sustained quantum oscillators and their emergent dynamics.

In this paper we study the quantum version of Stuart-Landau oscillators under attractive-repulsive diffusive coupling. We constitute the quantum master equation of the coupled system in the Lindblad form. Through a direct simulation of the quantum system we show a symmetry-breaking transition from quantum limit cycle oscillation to quantum inhomogeneous steady state or the quantum oscillation death state with increasing coupling strength. We also observe the same transition with the variation of Kerr parameter. This transition is different from the one reported in Refs.~\cite{qmod,qturing} where the symmetry-breaking transition occurs between the homogeneous steady state to inhomogeneous steady state. In the weak quantum regime, we support our results analytically using the noisy  classical model of the coupled system. We show that the symmetry-breaking scenario also holds good in the deep quantum regime.  Interestingly, we observe entanglement generation in the symmetry-breaking state that has no counterpart in the classical domain.

\section{Classical model: Attractive-Repulsive Diffusive Coupling}
We consider two identical classical Stuart-Landau oscillators, which are coupled via attractive-repulsive diffusive coupling scheme. The mathematical model is given by (written in a slightly modified form to make it compatible with the quantum version),
\begin{subequations}\label{eq:1}
\begin{align}
\dot x_j &= \omega y_j + \frac{k_1}{2} x_j - k_2({x_j}^2 + {y_j}^2)x_j \notag\\
&+ K({x_j}^2 + {y_j}^2)y_j - \epsilon ({x_j}^\prime - x_j),\\
\dot y_j &= -\omega x_j + \frac{k_1}{2} y_j - k_2({x_j}^2 + {y_j}^2)y_j \notag\\
&- K({x_j}^2 + {y_j}^2)x_j + \epsilon ({y_j}^\prime - y_j)
\end{align}
\end{subequations}
Here $j \in \{1,2\} $, $ \epsilon (>0)$ is the coupling parameter and $\omega$ is the common eigen frequency. The parameters $k_1$ controls the linear pumping, $k_2$ is the nonlinear damping parameter, and $K$ is the non-isochronous parameter. 
For $\epsilon=0$, (\ref{eq:1}) represents uncoupled nonisochronous Stuart-Landau oscillators .
The + (-) sign associated with the coupling parameter $\epsilon$ in Eq.~\ref{eq:1}(b) (Eq.~\ref{eq:1}(a)) determines the attractive (repulsive) coupling.
Equation \eqref{eq:1} can be represented in term of the amplitude $\alpha_j=x_j+i y_j$ that leads to the amplitude equation of the coupled system
\begin{equation}
\label{eq:2}
\dot \alpha_j = -i\omega \alpha_j + \frac{k_1}{2} \alpha_j - k_2{| \alpha_j |}^2 \alpha_j - i K {| \alpha_j |}^2 \alpha_j + \epsilon ( {\alpha_j}^\star - {\alpha_{j^\prime}}^\star).
\end{equation}

\begin{figure}
 \centering
 \includegraphics[width=.4\textwidth]{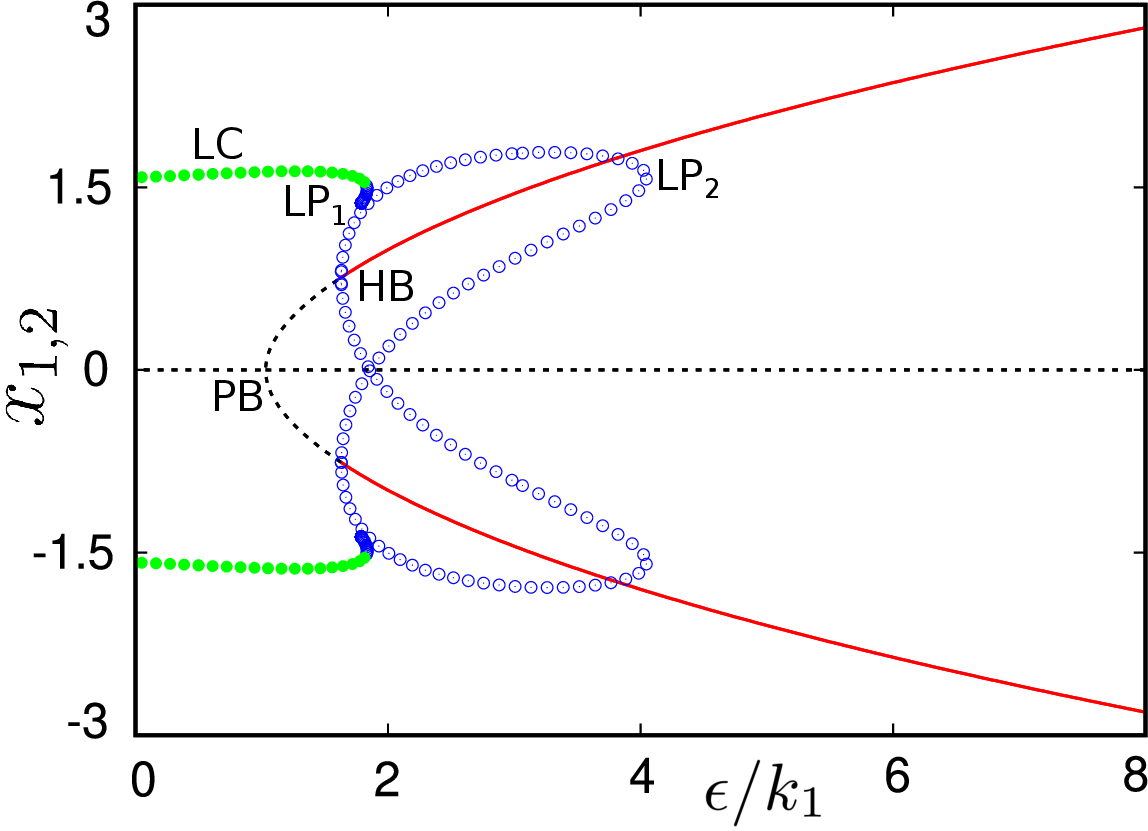}
 \caption{Bifurcation diagram of classical oscillators under attractive-repulsive diffusive coupling. Solid (dashed) line shows stable (unstable) steady state; Hollow circles represent unstable limit cycle; Filled circles denote stable limit cycle (LC). PB: pitchfork bifurcation, HB: Hopf bifurcation, LP$_{1,2}$: saddle node bifurcation of limit cycle. Parameters are $ \omega=2, k_1=1, k_2=0.2$ and $K=1$. }
 \label{fig:1}
\end{figure}

Equation (\ref{eq:1}) has a trivial homogeneous steady state (HSS) at the origin, $ F_{HSS} \equiv (0,0,0,0) $ and it has one additional non-trivial inhomogeneous steady state (IHSS) $ F_{IHSS} \equiv (x^\star, y^\star, -x^\star, -y^\star) $ where $ x^\star = \frac{-2 \epsilon \pm \sqrt{4 \epsilon^2-(\omega + \frac{k_1 K}{2 k_2})^2 + (\frac{2 \epsilon K}{k_2})^2}}{(\omega + \frac{k_1 K}{2 k_2}+\frac{2 \epsilon K}{k_2})} y^\star $. The Jacobian matrix of the system at the trivial fixed point is
$$\begin{pmatrix} \frac{k_1}{2}+\epsilon & \omega & -\epsilon & 0 \\ -\omega & \frac{k_1}{2}-\epsilon & 0 & \epsilon \\ -\epsilon & 0 & \frac{k_1}{2}+\epsilon & \omega \\ 0 & \epsilon & -\omega & \frac{k_1}{2}-\epsilon \end{pmatrix}.$$
The four eigenvalues of the system at the trivial fixed point (0,0,0,0) are
\begin{gather*}
\label{eq:eigenvalues}
\lambda_{1,2} = \frac{1}{2} [k_1 \pm 2 i \omega ], \\
\lambda_{3,4} = \frac{1}{2} [k_1 \pm 2 \sqrt{4{\epsilon}^2 - {\omega}^2}].
\end{gather*}
An eigenvalue analysis at the trivial fixed point shows that the system has one pitchfork bifurcation point at $ \epsilon_{PB}= \frac{1}{4} \sqrt{ {k_1}^2 + 4{\omega}^2}$ 
%Note that the occurrence of PB does not depend on the non-isochronous parameter $K$. 
where a symmetry-breaking bifurcation gives rise to unstable nontrivial fixed point $ F_{IHSS} \equiv (x^\star, y^\star, -x^\star, -y^\star)$, i.e, the inhomogeneous steady state (IHSS). 
Figure~\ref{fig:1} illustrates the bifurcation diagram of $x_1$ and $x_2$ with $\epsilon/k_1$ computed using XPPAUT \cite{xpp} (for $\omega=2, k_1=1, k_2=0.2 $ and $K=1$). 
The unstable $F_{IHSS}$ gets stability through a (subcritical) Hopf bifurcation (HB) at $\epsilon_{HB}/k_1\approx 1.63$. For a coupling strength less than $\epsilon_{LP_1}/k_1(\approx 1.83)$, a stable limit cycle exists, which is terminated by a saddle node bifurcation of limit cycle (denoted by LP$_1$). Therefore, the bifurcation diagram shows an abrupt transition from oscillatory state (LC) to stable inhomogeneous steady state, widely known as the oscillation death (OD) state \cite{kosprl}. In a narrow parameter zone lies between HB and LP$_1$ the system shows multistability consisting of steady state and limit cycles. Note that just  after the birth of the OD state, it coexists with an unstable limit cycle (shown by hollow circles in Fig.~\ref{fig:1}). However, for a greater value of coupling strength ($\epsilon$), two unstable limit cycles arising from two symmetric inhomogeneous branches collide with each other at $\epsilon/k_1\approx 4$ and annihilate through a saddle node bifurcation of limit cycle (LP$_2$), therefore, making the OD state the only existing dynamical state in the system.

\section{Quantum model and results}
Starting from the amplitude equation \eqref{eq:2}, we derive the quantum master equation \cite{chia,chia2023quantum} of attractive-repulsive duffusively coupled identical quantum Stuart-Landau oscillators that reads
%\begin{multline}
%\label{eq:3}
%\dot \rho = -i[ \omega (a_1^\dagger a_1 + a_2^\dagger a_2) + \frac{K}{2} ({a_1^\dagger}^2 a_1^2 + {a_2^\dagger}^2 a_2^2)\\- i \epsilon (a_1^\dagger a_2^\dagger - a_1 a_2) 
%+ \frac{ i \epsilon}{2} ({a_1^\dagger}^2 + {a_2^\dagger}^2 - a_1^2 - a_2^2),\rho]\\
%+ k_1 \sum_{j=1}^{2} \mathcal{D} [a_j^\dagger](\rho) + k_2 \sum_{j=1}^{2} \mathcal{D} [a_j^2] (\rho),
%\end{multline}
\begin{multline}
\label{eq:3}
\dot \rho = -i[(\mathcal{H}_0+\mathcal{H}_c),\rho]\\
+ k_1 \sum_{j=1}^{2} \mathcal{D} [a_j^\dagger](\rho) + k_2 \sum_{j=1}^{2} \mathcal{D} [a_j^2] (\rho),
\end{multline}
where $ a_j $ and $ a_j^\dagger $ are the annihilation and creation operator, respectively corresponding to the $j$-th oscillator ($j \in \{1,2\}$). $\mathcal{H}_0=\omega (a_1^\dagger a_1 + a_2^\dagger a_2) + \frac{K}{2} ({a_1^\dagger}^2 a_1^2 + {a_2^\dagger}^2 a_2^2)$ is the Hamiltonian of the uncoupled quantum Stuart-Landau oscillators in the presence of Kerr anharmonicity governed by the parameter $K$ \cite{brud-poch}. $\mathcal{H}_c=- i \epsilon (a_1^\dagger a_2^\dagger - a_1 a_2) + \frac{ i \epsilon}{2} ({a_1^\dagger}^2 + {a_2^\dagger}^2 - a_1^2 - a_2^2)$ is the coupling induced Hamiltonian arises due to the attractive-repulsive coupling. 
%%%%%%%%%%
The term $\mathcal{D}[\hat{L}](\rho)$ denotes the Lindblad dissipator: $\mathcal{D}[\hat{L}](\rho)=\hat{L}\rho \hat{L}^\dag-\frac{1}{2}\{\hat{L}^\dag \hat{L},\rho \}$, where $\hat{L}$ is an operator. As in the classical case, $k_1$ and $k_2$ are linear pumping rate and nonlinear damping rate, respectively: in the quantum domain $k_1$ controls the rate of one phonon creation and $k_2$ controls the rate of two-phonon absorption. Note that the attractive-repulsive coupling does not affect the dissipator of the quantum master equation.
%The presence of Kerr parameter ($K$) makes the system anharmonic \cite{brud-poch}. 
%%%%%%%%%%
In the classical limit $(k_1 \gg k_2)$ the pumping rate ($k_1$) dominates over the damping rate ($k_2$) and therefore a large number of phonons populate the excited energy levels making the system obeys classical probability distribution. Here we can take $\alpha=\langle a \rangle$ and using $\langle \dot a \rangle = Tr(\dot \rho a)$, the master equation (\ref{eq:3}) is equivalent to the amplitude equation of the classical system \eqref{eq:2}, which ensures that the master equation indeed represents the quantum version of the classical system.

\begin{figure}
 \centering
 \includegraphics[width=.45\textwidth]{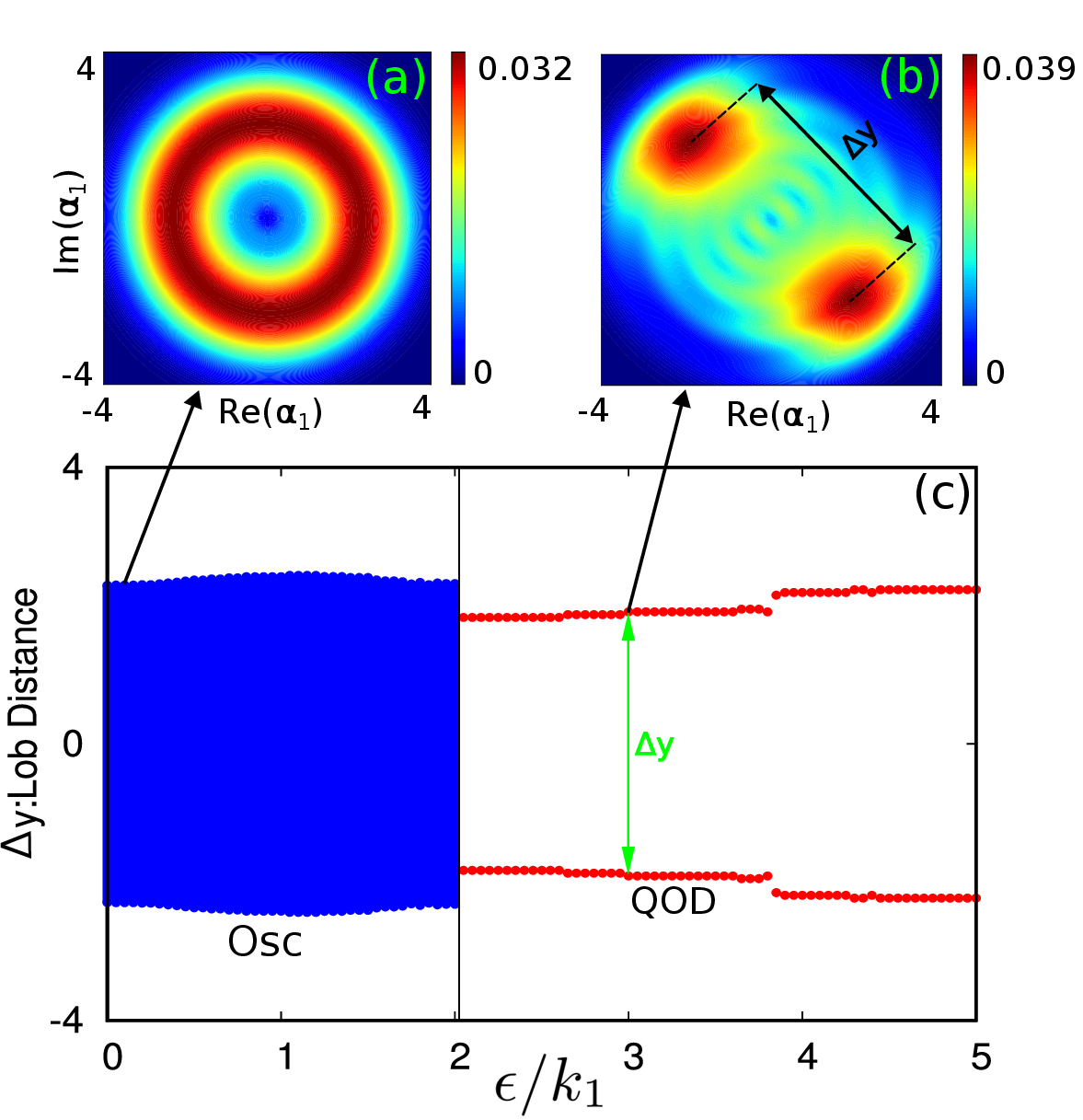}
 \caption{Wigner function representation of (a) the quantum limit cycle for $ \epsilon / k_1 = 0.1$ and (b) quantum OD state for $ \epsilon / k_1 = 3 $. (c) The variation of the distance between local maximum values of the Wigner function or lobe distance ($\Delta y$) plotted with $\epsilon/k_1$ that shows a transition from quantum limit cycle to QOD. The oscillatory zone (Osc) is indicated with the filled area containing the points distributed along the periphery of the ring of the maximum Wigner function distribution of the QLC. Other parameters are $ \omega=2, k_1=1, k_2=0.2$ and $K=1$.}
 \label{fig:lobdistance_dif}
\end{figure}

To explore the dynamics of the quantum system, we carry out a direct simulation of the quantum master equation (\ref{eq:3}) (using QuTiP \cite{qutip,qutip1}) and track the system dynamics in the phase space using Wigner function \cite{carmichael}. 
We observe  a quantum limit cycle (QLC) at a lower coupling strength ($\epsilon/k_1$); Fig.~\ref{fig:lobdistance_dif}(a) shows it at an exemplary value $\epsilon / k_1=0.1$ ($ \omega=2, k_1=1, k_2=0.2$ and $K=1$).  With an increasing coupling strength, the continuous rotational symmetry is broken and the quantum limit cycle is converted into an inhomogeneous steady state, namely the quantum oscillation death (QOD) characterized by the appearance of two-lobed Wigner function distribution; Fig.~\ref{fig:lobdistance_dif}(b) illustrates a QOD state with two prominent lobes separated by a distance $\Delta y$. 
To show the transition from oscillation to QOD is continuous we track the lobe distance $\Delta y$, which is the distance between the local maxima of the Wigner distribution function. Figure~(\ref{fig:lobdistance_dif})(c) demonstrates the variation of $\Delta y$ with increasing $\epsilon$ (with $k_1=1$). 
The oscillatory region (indicated by Osc) is indicated by the filled area containing the points distributed along the periphery of the ring of the maximum Wigner function distribution of the QLC. Beyond a certain value of $ \epsilon $, the QOD state appears, which is indicated by a non-zero $\Delta y$, where the Wigner function splits into a bimodal shape whose lobes are separated by $\Delta y$. 
%The symmetry-breaking transition with increasing $\epsilon/k_1$ ($k_1=1$) from quantum limit cycle to quantum oscillation death state is shown in Figure \ref{fig:lobdistance_dif} (c). 
Note that the present symmetry-breaking transition scenario differs from the one reported in previous studies (see \cite{qmod,qturing}), where the symmetry-breaking transition was found from quantum homogeneous steady state (namely, quantum amplitude death state) to quantum OD state. In our system, no quantum amplitude death is observed. Further, unlike the classical case, interestingly, we have not observed any multistability in the corresponding quantum phase transition. This again confirms the fundamental difference between a classical and quantum system. The absence of hysteresis and bistability in the quantum system may be ascribed to the inherent quantum noise, which washed out any initial condition dependent states \cite{chia,brud-poch}.

\begin{figure}
 \centering
 \includegraphics[width=.45\textwidth]{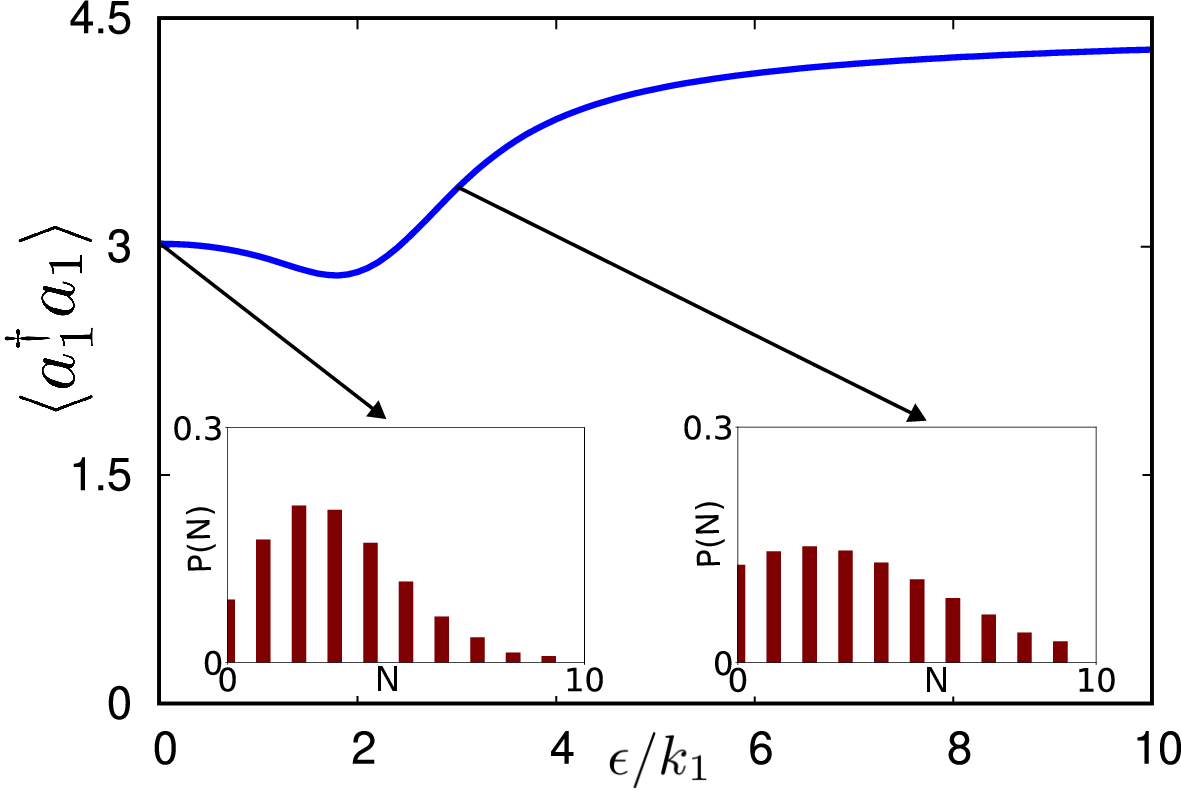}
 \caption{The steady state mean phonon number $ \langle a_1^\dagger a_1 \rangle (= \langle a_2^\dagger a_2 \rangle ) $ with $\epsilon / k_1$. Inset shows the Fock distribution function in the oscillatory state for (a) $ \epsilon / k_1 = 0.01$ and quantum OD state for (b) $ \epsilon / k_1 = 3 $. Other parameters are $ \omega=2, k_1=1, k_2=0.2$ and $ K=1 $. }
 \label{fig:3}
\end{figure}

Next, we compute the steady state mean phonon number $ \langle a_1^\dagger a_1 \rangle (= \langle a_2^\dagger a_2 \rangle ) $ with $ \epsilon/k_1 $. Figure~\ref{fig:3} demonstrates the variation of mean phonon number $\langle a_1^\dagger a_1 \rangle$ with $ \epsilon/k_1$ ($k_1=1$). It shows an initial dip in the mean phonon number and then in the QOD region rises to a certain value and remains constant there. 
Note that the mean phonon number does not reduce to a near-zero value confirming the absence of quantum amplitude death state \cite{qad1}. 
The variation of the mean phonon number is also verified by inspecting the Fock distribution function. The insets of Fig.~\ref{fig:3} show the occupation probability in Fock states in the oscillatory zone (for $\epsilon / k_1 = 0.01$) and quantum OD state (for $\epsilon / k_1 = 3$). 
%Note that in the QOD state, the Fock distribution function is much more dispersed compared to that of oscillatory state.
For a limit cycle, Fock levels show a prominent peak far from ground state and decays in both side. In the QOD state, more phonons start coming near the ground state, thus Fock levels around the ground state shows relatively an increased occupancy. However, in the QOD state the average mean phonon number is not zero but shows a considerable nonzero value, therefore, unlike limit cycle, here the Fock distribution is reorganized showing a less sharp peak and a relatively wide-spread distribution.

\begin{figure}
 \centering
 \includegraphics[width=.45\textwidth]{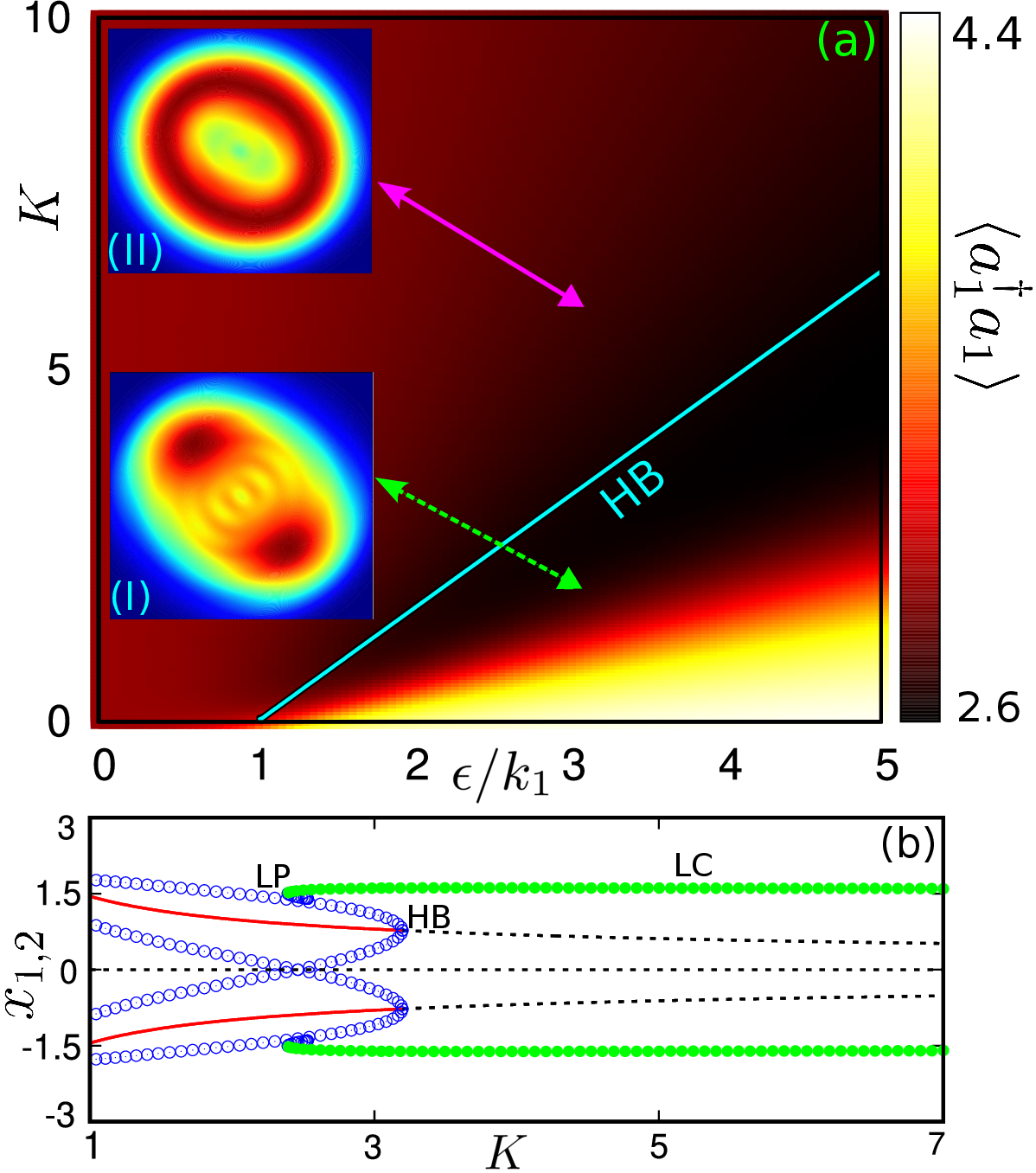}
 \caption{(a) Two parameter diagram in the $ \epsilon/k_1 - K $ space showing the mean phonon number $\langle a_1^\dagger a_1 \rangle (= \langle a_2^\dagger a_2 \rangle)$ of both the oscillators. The solid line represents Hopf bifurcation (HB) in the classical oscillators that differentiates between the oscillatory and OD region. The inset (I) shows the quantum OD state at $K= 2$ and (II) shows the quantum limit cycle at $K=6$ (both for $\epsilon/k_1 = 3$). (b) Classical bifurcation diagram (using Eq.~\eqref{eq:1}) with the variation of $K$ (at $\epsilon/k_1 = 3$): it shows a transition from OD to limit cycle (LC) state with increasing $K$. Other attributes are same as Fig.~\ref{fig:1}. Parameters are $ \omega=2, k_1=1, k_2=0.2$.}
 \label{fig:twopar}
\end{figure}

Finally, we investigate the effect of Kerr parameter ($K$) on the symmetry-breaking transition. Interestingly, we observed that an increasing Kerr parameter gives rise to an (inverse) symmetry-breaking transition from QOD to quantum limit cycle. We support our observation in Fig.~\ref{fig:twopar}(a) showing the color map of the mean phonon number in the $\epsilon/k_1-K$ space. The solid line indicates the classical Hopf bifurcation (HB) curve.
%; it indicates that for higher value of $K$, a stronger coupling strength is required to induce symmetry-breaking transition. 
The inset (I) of the figure shows the quantum oscillation death for $ K=2 $ and inset (II) shows the quantum oscillatory state for $ K=6 $ (for fixed $ \epsilon/k_1=3 $). Therefore, we can conclude that an increasing Kerr parameter for a fixed $ \epsilon/k_1 $ value shows a transition from QOD to quantum oscillation, which is also contrary to the previously known result reported in Ref.~\cite{qkerr} that an increasing Kerr nonlinearity gives a transition from QOD to quantum amplitude death state.
The transition from QOD to quantum limit cycle is supported by the corresponding classical bifurcation diagram (Fig.~\ref{fig:twopar}(b)): it shows a complex bifurcation structure where the OD state is transformed into a stable limit cycle via Hopf bifurcation (HB) with increasing Kerr parameter. The stable limit cycle is terminated through a saddle node bifurcation of limit cycle (denoted by LP) for a decreasing $K$. Like Fig.~\ref{fig:1}, here also the bifurcation structure shows a multistable region between HB and LP, however, due to inherent quantum noise no multistabilty is observed in the quantum regime.

\section{Analytical treatment: Noisy classical model}
Next we compare the results of quantum dynamics with the corresponding noisy classical model. 
%where the classical dynamics is considered in presence of a finite amount of noise with the equal intensity of quantum noise and we need to solve a stochastic differential equation. 
For this, the quantum master equation (\ref{eq:3}) is represented in phase space using partial differential equation of the Wigner distribution function $ W(\alpha)$ (\cite{carmichael}):
\begin{multline}
\label{eq:4}
{\partial}_t W(\alpha,t) = \sum_{j=1}^{2} \Biggl[-\Biggl(\frac{\partial}{\partial {\alpha}_j} {\mu}_{\alpha_j} + c.c.\Biggl) \\
 + \frac{1}{2}\Biggl(\frac{\partial^2}{ \partial\alpha_j \partial{\alpha_j}^* } D_{ \alpha_j {\alpha_j}^* } + \frac{\partial^2}{ \partial\alpha_j \partial{\alpha_{j^\prime}}^* } D_{\alpha_j {\alpha_{j^\prime}}^* }\Biggl) \\
 + (-\frac{iK}{4} + \frac{k_2}{4}) \Biggl(\frac{\partial^3}{ {\partial {\alpha_j}^*} {\partial{\alpha_j}^2} } \alpha_j + c.c. \Biggl)\Biggl] W(\alpha ,t),
\end{multline}
where
\begin{multline*}
\begin{split}
\mu_{\alpha_j} &= \Biggl[ -i\omega + \frac{k_1}{2} - k_2( {|\alpha_j|}^2 - 1) - iK({|\alpha_j|}^2 - 1) \Biggl] \alpha_j \\
&+ \epsilon (\alpha_j^\star - {\alpha_{j^\prime}}^*),
\end{split}
\end{multline*}
$$ D_{\alpha_j {\alpha_j}^* } = k_1 + 2 k_2 (2 {|\alpha_j|}^2 - 1), $$
$$ D_{\alpha_j {\alpha_{j^\prime}}^* } = 0, $$
with $j$ and $ j^\prime = 1,2 $ and $ j \neq j^\prime$. Here $\mu_{\alpha}$ and $D_{\alpha {\alpha}^* }$ are the components of drift and diffusion vector, respectively. In the weak quantum regime where the linear pumping is dominant (i.e., $ k_1>k_2 $ ), the coefficient of the third-order derivative term becomes smaller, therefore, we can neglect this term in Eq.~\ref{eq:4} \cite{qad1}: thus, in this approximation, Eq.~\ref{eq:4} reduces to the following Fokker-Planck equation \cite{carmichael},
\begin{multline}
\label{eq:5}
{\partial}_t W(\textbf{X}) = \sum_{j=1}^{2} \Biggl[ - \Bigl(\frac{\partial}{\partial x_j} \mu_{x_j} + \frac{\partial}{\partial y_j} \mu_{y_j} \Bigl) \\
+ \frac{1}{2} \Bigl(\frac{\partial^2}{\partial x_j \partial x_j} D_{x_j x_j} + \frac{\partial^2}{\partial y_j \partial y_j} D_{y_j y_j} \\
+ \frac{\partial^2}{\partial x_j \partial x_{j^\prime}} D_{x_j x_{j^\prime}} + \frac{\partial^2}{\partial y_j \partial y_{j^\prime}} D_{y_j y_{j^\prime}}\Bigl)\Biggl] W(\textbf{X}),
\end{multline} 
where, $\textbf{X} = (x_1,y_1,x_2,y_2)$ and the components of drift vector are
\begin{multline}
\begin{split}
\label{eq:6}
\mu_{x_j} &= \Bigl[ \omega + K({x_j}^2+{y_j}^2-1) \Bigl] y_j\\
&+ \Bigl[\frac{k_1}{2} - k_2({x_j}^2+ {y_j}^2 - 1) + \epsilon\Bigl] x_j - \epsilon x_{j^\prime},\\
\mu_{y_j} &= \Bigl[ - \omega - K({x_j}^2+{y_j}^2-1) \Bigl] x_j\\
&+ \Bigl[\frac{k_1}{2} - k_2({x_j}^2+ {y_j}^2 - 1) - \epsilon\Bigl] y_j + \epsilon y_{j^\prime}.
\end{split}
\end{multline}
The corresponding diffusion matrix can be written in the following form:
\begin{equation}
\label{eq:7}
\mathbf{D} = \frac{1}{2}
\begin{pmatrix}
\nu_1 & 0 & 0 & 0 \\
0 & \nu_1 & 0 & 0 \\
0 & 0 & \nu_2 & 0 \\
0 & 0 & 0 & \nu_2 
\end{pmatrix},
\end{equation}
where $ \nu_j = \frac{k_1}{2} + k_2[2({x_j}^2 + {y_j}^2)-1]$.

\begin{figure}
 \centering
 \includegraphics[width=.45\textwidth]{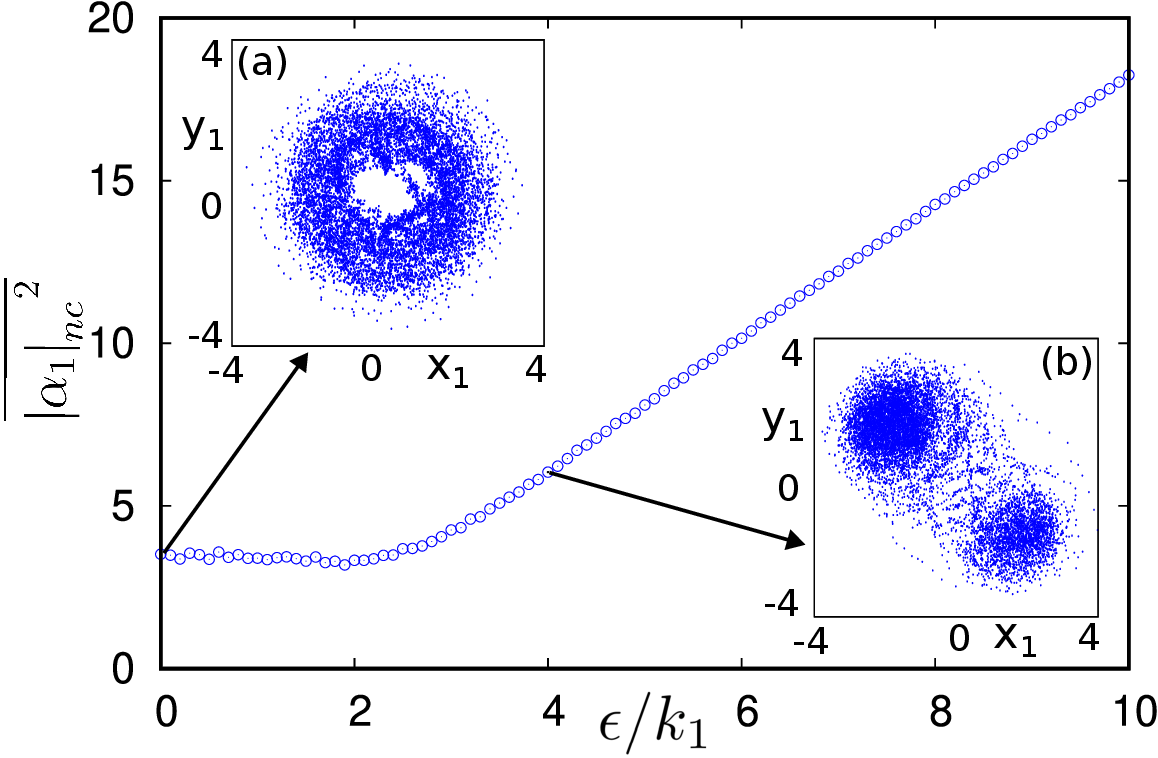}
 \caption{Variation of the averaged amplitude from the noisy classical model $(\overline{{{|\alpha_1|}_{nc}}^2})$ of the first oscillator with coupling strength. Insets show the phase space plot of (a) noisy limit cycle (for $ \epsilon=0.01$), and (b) noisy QOD (for $ \epsilon=4.0$). Other parameters are $ \omega=2, k_1=1, k_2=0.2 $ and $ K=1 $.}
 \label{f:noisy}
\end{figure}

From \eqref{eq:5} we get the following stochastic differential equation:
\begin{equation}
\label{eq:8}
d\textbf{X} = \boldsymbol{\mu} dt + \boldsymbol{\sigma} d {\textbf{W}}_t,
\end{equation}  
where $\boldsymbol{\mu}$ is the drift vector, $\sigma$ is the noise strength and $d{\textbf{W}}_t$ is the Wigner increment. As the diffusion matrix $\textbf{D}$ is diagonal, we can derive noise strength $\boldsymbol{\sigma}$ from the diffusion matrix as $\boldsymbol{\sigma}= \sqrt \textbf{D}.$
Thus the $ \boldsymbol{\sigma} $ matrix can be written as
\begin{equation}
\label{eq:9}
\boldsymbol{\sigma} =
\begin{pmatrix}
\sqrt{\frac{\nu_1}{2}} & 0 & 0 & 0 \\
0 & \sqrt{\frac{\nu_1}{2}} & 0 & 0 \\
0 & 0 & \sqrt{\frac{\nu_2}{2}} & 0 \\
0 & 0 & 0 & \sqrt{\frac{\nu_2}{2}}
\end{pmatrix}.
\end{equation}
We solve \eqref{eq:8} numerically (using python JiTCSDE module \cite{jitcode}) and compute the ensemble average of squared steady-state amplitude of the oscillator. The variation of the average amplitude from the noisy classical model $ (\overline{{{|\alpha_1|}_{nc}}^2}) $ with $\epsilon/k_1$ is shown in Fig.~\ref{f:noisy}: with increasing coupling strength, it shows an initial dip and then an increasing nature that is in accordance with the quantum model. In the quantum regime (Fig.~\ref{fig:3}) the maximum value of the average phonon number is limited by the finite dimension of the Hilbert space, however, no such restriction on the average amplitude exists in the noisy classical case (Fig.~\ref{f:noisy}). Nevertheless, the results of both quantum and noisy-classical cases agree well in the limit of small and moderate coupling strength.
Figure~\ref{f:noisy}(a) shows the noisy limit cycle in phase space at an exemplary value $\epsilon/k_1=0.01$. An increase in $(\overline{{{|\alpha_1|}_{nc}}^2})$ beyond $\epsilon/k_1\approx 2$ indicates the appearance of symmetry-breaking state; this is shown in phase space in inset (b) for an exemplary value $\epsilon/k_1=4$. It can be seen that unlike noisy limit cycle, here in the symmetry-breaking state the density of phase points are higher in two regions of phase space indicating the noisy version of QOD.

\begin{figure}
 \centering
 \includegraphics[width=.45\textwidth]{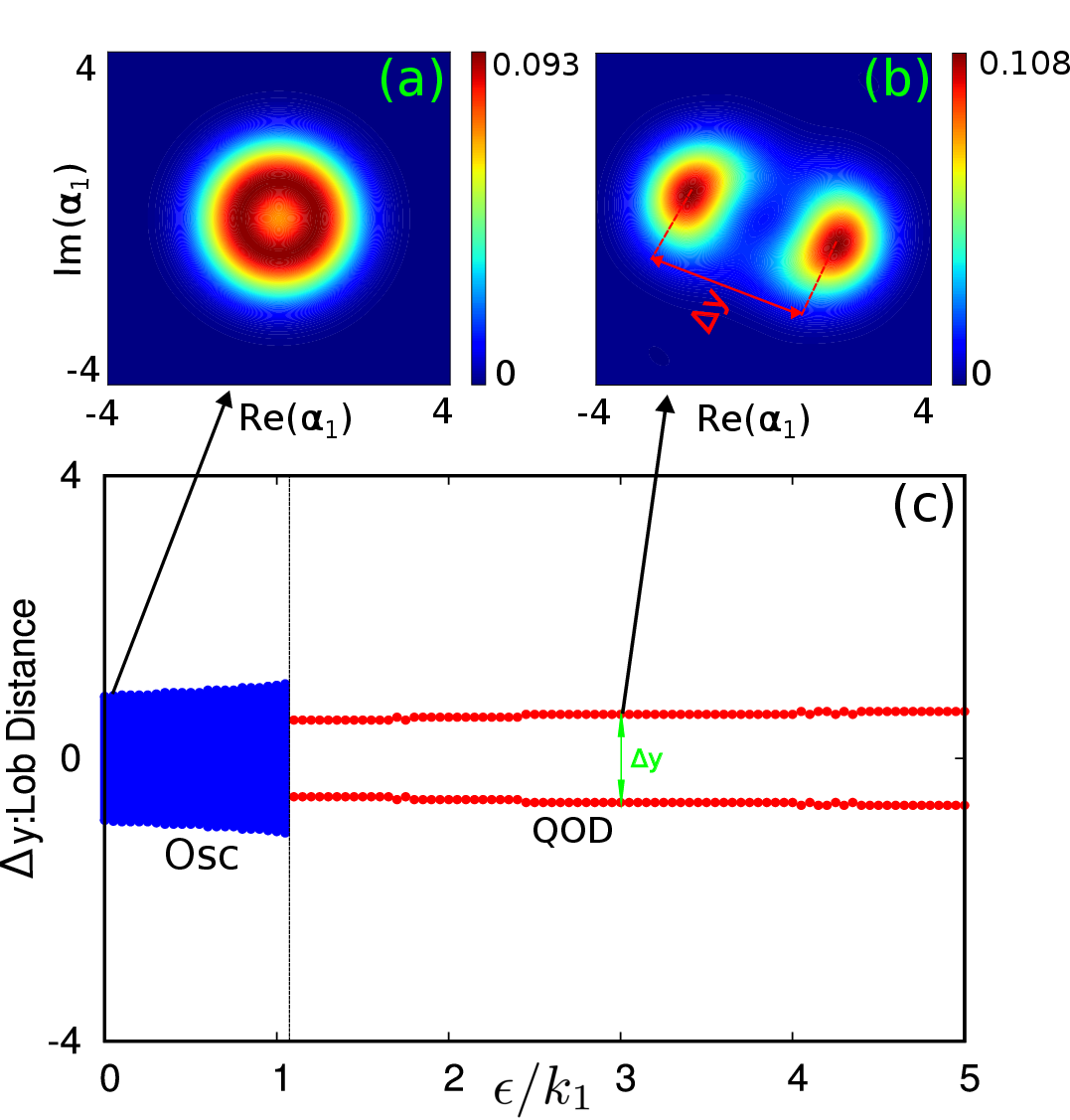}
 \caption{Deep quantum regime ($k_2>k_1$): (a) Quantum limit cycle at $\epsilon/k_1=0.01$ (b) QOD at $\epsilon/k_1=3$. (c) Lobe distance ($\Delta y$) with $\epsilon/k_1$; it shows a transition from quantum limit cycle oscillation (Osc) to quantum oscillation death (QOD). Other parameters are $\omega=2, k_1=1, k_2=3$, $K=1$.}
 \label{fig:deep}
\end{figure}
\section{Deep quantum regime}

We also explore the symmetry-breaking scenario in the deep quantum regime. This regime is characterized by high damping, i.e., $k_2>k_1$; since Fock levels near the ground state are more populated states, quantum noise plays a crucial role in shaping the dynamics in this regime. We take $k_1=1$ and $k_2=3$ and simulate the dynamics of \eqref{eq:3}. Here also we observed a transition from quantum limit cycle to quantum OD state: Figure~\ref{fig:deep}(a) shows the illustrative example of the quantum limit cycle for $\epsilon/k_1=0.01$, and Fig.~\ref{fig:deep}(b) shows the same of quantum OD state for $\epsilon/k_1=3$. The continuous transition from quantum LC to QOD is tracked by the lobe distance between the symmetry-breaking inhomogeneous states and is shown in Fig.~\ref{fig:deep}(c). It shows a sudden transition from the oscillatory state to inhomogeneous steady state at $\epsilon/k_1\approx 1.05$. Note that for the same $\epsilon/k_1$ value, the amplitude of the limit cycle and the lobe distance of the symmetry-breaking states are lesser in comparison with the weak quantum regime (cf. Fig.~\ref{fig:lobdistance_dif}(c)), which is the characteristic of the deep quantum regime. 

\begin{figure}
 \centering
 \includegraphics[width=.32\textwidth]{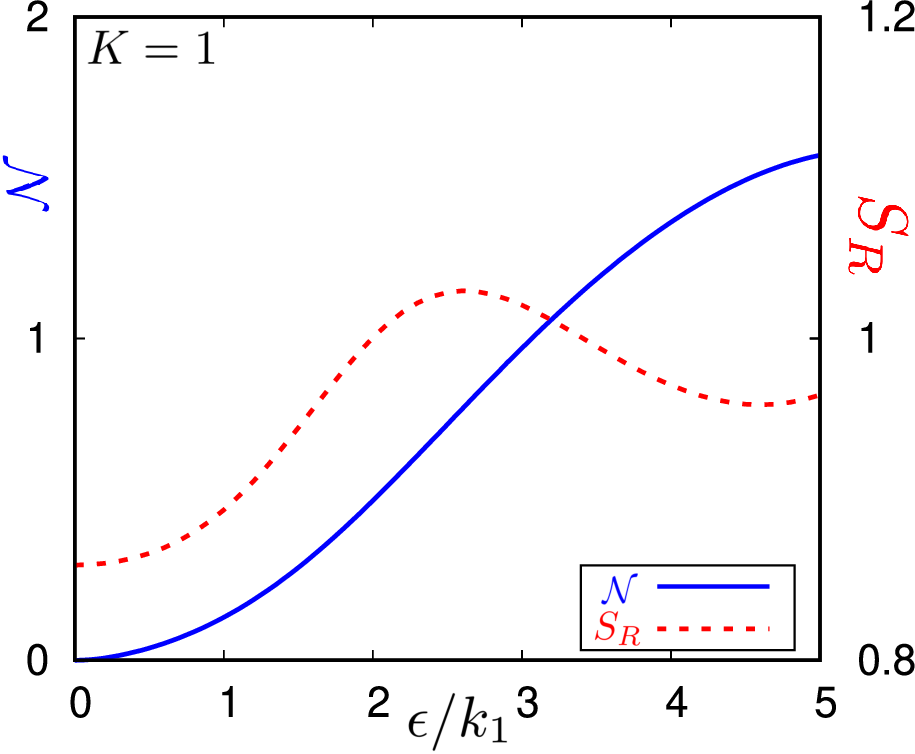}
 \caption{The plot of negativity ( $ \mathcal{N} $ ) and second-order R\'enyi entropy $(S_R)$ with $\epsilon/k_1$. Other parameters are $\omega=2, k_1=1, k_2=3$, $ K = 1 $.}
 \label{fig:neg}
\end{figure}

Next, we investigate the generation of entanglement, which is a direct consequence of quantum correlation and mixedness in the coupled system. It has been shown in Ref.~\cite{kato} that the transition from homogeneous steady state to inhomogeneous steady state is associated with generation of quantum mechanical entanglement. Here we explore whether or not quantum entanglement may be produced during the transition from an oscillatory state to an inhomogeneous steady state. For this we employ the Negativity parameter that is a measure of quantum entanglement defined as $ \mathcal{N} = ({||{\rho}^{\Gamma_1}||}_1 - 1 )/2 $, where $ {\rho}^{\Gamma_1} $ is the partial transpose of the density matrix $ \rho $ of the system with respect to first oscillator and $ {||X||}_1 = \mathrm{Tr} |X| = \mathrm{Tr} \sqrt{{X}^\dagger X} $ (here we have used $X = {\rho}^{\Gamma_1}$). The value of $ \mathcal{N} = 0 $ indicates the absence of entanglement, and as $ \mathcal{N} $ increases entanglement increases. 
%Ref. [\ref{ref:1}] established that the enhancement of $ \mathcal{N} $ for the higher values of $ \epsilon $ and lower values of $ K $ , as the coupling strength increases the entanglement increases and also the Kerr nonlinearity does not supports the symmetry-breaking state. 
In Fig.~\ref{fig:neg} we observe that in the uncoupled condition $(\epsilon/k_1=0)$ the value of the negativity parameter is zero, indicating the absence of entanglement between the two oscillators, however, as the coupling strength increases the corresponding $ \mathcal{N}$ also rises to a non zero value indicating the generation of the quantum mechanical entanglement. Therefore, we conclude that although, the transition scenarios of this paper and Ref.~\cite{kato} are different, however, the underlying mechanism of increased entanglement remains the same in both the cases.

In the context of entanglement and quantum correlations another reliable measure is  R\'enyi entanglement entropy. The R\'enyi entropy of order $\alpha$ for a density matrix $\rho$ is defined as \cite{qchm,renyi-science}: $S_{\alpha}(\rho) = \frac{1}{1-\alpha } \log (Tr(\rho^{\alpha}))$, where $\alpha$ is a positive real number and $Tr(\rho^{\alpha})$ is the trace of the $\alpha$-th power of the density matrix. 
%The choice of parameter $\alpha$ gives different types of entropy; for example, for $\alpha=1$, R\'enyi rentropy reduces to the von Neumann entropy, whereas 
The second-order R\'enyi entropy of the reduced matrix ($\alpha=2: S_{2}(\rho)=S_R$) is used to probe the quantum entanglement. $S_R$ with $\epsilon/k_1$ (Fig.~\ref{fig:neg}) shows an increasing entanglement in the transition from quantum limit cycle to QOD sate (note that, unlike  $\mathcal{N}$, $S_R$ does not start from zero since uncoupled limit cycles are not a pure state \cite{renyi-science}). The R\'{e}nyi entropy reveals an additional information about the degree of mixedness: it shows that with increasing coupling, $S_R$ reaches a maximum value and then at higher coupling regime, it decreases and settles at a lower value (that is still greater than the uncoupled value of $S_R$ at $\epsilon=0$). This may be attributed to the fact that a stronger coupling leads to an increased dephasing \cite{squeezing} that in turn reduces the quantum mechanical correlation between its units.

%##############################################################
%###################### Conclusions ###########################
%##############################################################
\section{Conclusions}
\label{sec_conc}
In this paper we have explored the role of attractive-repulsive coupling in shaping the collective behavior of coupled oscillators in the quantum domain. A direct simulation of the quantum master equation of the coupled quantum Stuart-Landau oscillators showed a symmetry-breaking transition from quantum limit cycle to quantum oscillation death state with increasing coupling strength or decreasing Kerr nonlinearity parameter. This is in contrast to the quantum symmetry-breaking transitions reported earlier where inhomogeneity emerges from the homogeneous steady state \cite{qmod,qturing,qkerr}.
The phenomenon is general as it occurs at both weak and deep quantum regime. Specially, in the deep quantum regime where quantum noise is strong, yet it can not wash out the inhomogeneous state indicating that the symmetry-breaking state is indeed a prominent state.
%This is in contrast to the quantum symmetry-breaking states reported earlier where homogeneity comes from the homogeneous steady state \cite{qmod,qtur}.
An analysis of the noisy classical model supports the observed results.  
A positive correlation of the symmetry-breaking state to the quantum mechanical entanglement generation has also been established that has no classical counterpart.

As mentioned, unlike previous studies on symmetry-breaking transitions, the attractive-repulsive coupling does not show a homogeneous steady state (or quantum amplitude death), rather it gives a direct transition from quantum limit cycle to quantum OD. This can be intuitively explained from the form of the quantum master equation \eqref{eq:3}: Note that the master equation does not contain a term corresponding to single phonon loss (i.e., $\mathcal{D} [a_j] (\rho)$ term), which is essential for relaxing the system to the ground state and thus generating a quantum amplitude death state. This makes the attractive-repulsive coupling unique from other coupling schemes.
%application
We believe that the present coupled system can be engineered in experimental set up with quantum electrodynamics \cite{qed-expt-two-photon-loss,qed-expt}, trapped ion \cite{expt-ion} and optomechanical experiment \cite{expt-mem}.   
The present study will deepen our insight of strongly interacting dissipative quantum systems that has immense applications in quantum science and engineering \cite{photon1,photon2}. 

\begin{acknowledgments}
B.P. acknowledges the financial assistance from DST-INSPIRE. T. B. acknowledges the financial support from the Science and Engineering Research Board (SERB), Government of India, in the form of MATRICS research Grant [MTR/2022/000179].
\end{acknowledgments}

%\bibliography{qattrep}
\providecommand{\noopsort}[1]{}\providecommand{\singleletter}[1]{#1}%
\end{document}